\documentclass[conference]{IEEEtran}
\IEEEoverridecommandlockouts
\usepackage{cite}
\usepackage{amsmath,amssymb,amsfonts}
\usepackage{algorithmic}
\usepackage{graphicx}
\usepackage{textcomp}
\def\BibTeX{{\rm B\kern-.05em{\sc i\kern-.025em b}\kern-.08em
    T\kern-.1667em\lower.7ex\hbox{E}\kern-.125emX}}

\let\IEEEproof\proof
\let\endIEEEproof\endproof
\let\proof\relax
\let\endproof\relax
\usepackage{amsthm}
\let\proof\IEEEproof
\let\endproof\endIEEEproof

\usepackage{hyperref}
\usepackage{csquotes}
\newtheorem{theorem}{Theorem}[section]

\newtheorem{proposition}[theorem]{Proposition}

\theoremstyle{definition}

\theoremstyle{remark}

\DeclareMathOperator*{\argmin}{arg\,min}
\DeclareMathOperator*{\argmax}{arg\,max}

\usepackage[dvipsnames]{xcolor}
\usepackage{tikz}
\usepackage{booktabs}

\usepackage[caption=false,font=footnotesize]{subfig}
\usepackage{graphicx} 

\usetikzlibrary{shapes.geometric}
\definecolor{mplorange}{HTML}{FFA500}
\definecolor{viridisdark}{HTML}{440154}

\usepackage{booktabs}
\usepackage{pgfplotstable}
\pgfplotsset{compat=1.18}

\usepackage{footmisc}

\newcommand{\symKnown}{\tikz[baseline=-0.6ex]\fill[magenta, opacity=0.85] (0,0) circle (0.875pt);}

\newcommand{\symPast}{\tikz[baseline=-0.6ex]{%
  \fill[viridisdark] (-1pt,-2pt) rectangle (9pt,2pt);
  \draw[white, opacity=0.8, line width=0.4pt] (0,0) -- (8pt,0);
  \fill[white, opacity=0.8] (4pt,0) circle (0.75pt);}}

\newcommand{\symPlanned}{\tikz[baseline=-0.6ex]{%
  \draw[cyan, line width=0.5pt, dash pattern=on 1.85pt off 0.8pt] (0,0) -- (8pt,0);
  \fill[cyan] (4pt,0) circle (0.75pt);}}

\newcommand{\symCurrent}{\tikz[baseline=-0.6ex]\draw[black, fill=red, line width=0.5pt] (0,0) circle (2.25pt);}

\newcommand{\symTarget}{\tikz[baseline=-0.6ex]\node[star, star points=5, star point ratio=2.618,
  minimum size=7pt, inner sep=0pt, draw=black, fill=mplorange, line width=0.5pt] {};}

\begin{document}

\title{Active Learning of Flows from Curve Observations using Gaussian Processes\\
\thanks{This work was supported by the Deutsche Forschungsgemeinschaft (DFG, German Research Foundation) within the Research Unit ALeSCo—Active Learning in Systems and Control under Project 535860958 and within the RTG METEOR – GRK 3081/1 – Project number 534429653.}
}

\author{\authorblockN{Konrad Reichel}
\authorblockA{\textit{Chair of Information-Oriented Control}\\
\textit{Technical University of Munich}\\
Munich, Germany\\
konrad.reichel@tum.de}
\and
\authorblockN{Sandra Hirche}
\authorblockA{\textit{Chair of Information-Oriented Control}\\
\textit{Technical University of Munich}\\
Munich, Germany\\
hirche@tum.de}
}

\maketitle

\begin{abstract}
The active learning of real control systems requires informative roll-outs of controls. In this paper, we present a local optimization algorithm that finds the trajectory that most reduces uncertainty about the flow of a control system modeled by a Gaussian process. Therefore, we derive an approximation scheme for the posterior kernel of the Gaussian process after observing it on an entire continuous-time trajectory and establish a Pontryagin-type necessary condition for the optimizer.
As the flow is unknown, the actual trajectory can deviate from the planned trajectory. We propose a trajectory planner that searches for the most informative control within a maximal allowed deviation horizon instead of a fixed time horizon. We show by trajectory-wise time rescaling that the underlying optimal control problem is equivalent to a fixed end time optimal control problem.
Finally, we study both the fixed end time planner and the fixed maximum deviation planner in a numerical example and compare them with benchmarks.
\end{abstract}


\section{Introduction}
Active learning describes the intentional excitation of an unknown system to identify its dynamics. The fundamental difficulty of this control task is that trajectories have to be planned without having a model, but for modeling, the system has to be steered into regions with high uncertainty about its dynamics. \\
In this paper, we focus on methods that model the vector field of the control system or parts of it with a Gaussian process. Gaussian processes comprise a genuine uncertainty measure by means of their variance at a certain point. The correlation between the vector fields at multiple points admit inference about the vector field at unseen points. Furthermore, prior knowledge in form of simple models or data points can be incorporated easily. \\
Most active learning methods try to steer the system directly into regions with high uncertainty. Some pick the point in the joint space and control space whose dynamics have the highest variance and roll out a trajectory to that point~\cite{pmlr-v120-buisson-fenet20a}. The algorithm in~\cite{algp_local_global} explores the neighborhood of that point afterward. As the roll-out of trajectories might be costly, multi-step methods consider the variance along the entire path~\cite{pmlr-v120-buisson-fenet20a, algp_local_global, rhal_Le}. Multi-step planners also perform better than greedy one step planners\cite{pmlr-v120-buisson-fenet20a}. 
Points with the highest variance must not be the ones containing most information about the other points in the domain of the Gaussian process; e.g., for radial basis function kernels, these approaches tend to measure the Gaussian process at the boundaries of the domain which is suboptimal~\cite{JMLR:v9:krause08a}. 
In contrast, other approaches attempt to identify the points that contain much information about the Gaussian process across the entire domain. In a model predictive control fashion, \cite{pmlr-v120-capone20a} approaches the point that contains most information about the dynamics and updates the Gaussian process after each step. 
In \cite{taskspecific_al} the trajectories are planned such that the posterior variance at points on a trajectory that should be tracked as a downstream task is minimized. The underlying optimal control problem is solved by a gradient-free optimizer. \\
All of these approaches use GP mean-based trajectory planners but neglect deviations between true and nominal trajectory. Consequently, it can happen that the planner assumes it can pass informative locations that are not reachable. Moreover, the underlying optimal control problems use fixed time horizons. In scenarios where parts of the region reachable within the planning horizon remain unidentified, that seems sufficient for planning meaningful trajectories. However, inherently adaptive methods that extend the horizon when the neighborhood has already been identified are lacking. Furthermore, to the best of our knowledge, there is as yet no active learning method that plans trajectories to minimize the posterior variance across the entire state space. 
Lastly, all approaches above only consider the observation of finitely many data points, although they assume access to measurements at any time and adding more, i.e. up to infinite data, would decrease the variance further~\cite[Thm. 2.6.5]{cover2005elements}. \\
The contributions of this paper are two active learning methods for flow identification. Both plan trajectories to minimize the posterior variance across the entire state space while conditioning the Gaussian process on the entire continuous-time trajectory. We derive the gradient of this objective w.r.t. the control which facilitates gradient-based optimization and a Pontryagin-type optimality condition. 
In addition, the second method implements an optimal control problem that restricts its search space to trajectories for which the deviation between true and planned trajectory can be guaranteed to remain within a predefined bound with high probability. That results in an adaptive planning horizon depending on the faced uncertainty. We show by trajectory-wise time rescaling that this problem is equivalent to a fixed end time optimal control problem with fewer side constraints and optimization variables. 
A numerical example shows that our methods can outperform the existing state of the art i.e., $\textit{rec}$ and $p\&a$ from~\cite{pmlr-v120-buisson-fenet20a}, in terms of root mean square error reduction.

\section{Problem Statement}
The ground truth control system is assumed to be
\begin{equation} 
    \dot{x}(t) = f(x(t), u(t)) + r(x(t)), \ x(0) = x_0,
    \label{eq:ground_truth_control_system} 
\end{equation}
with state $x: \mathbb{R}_{\geq 0} \!\to\! \mathcal{X} \!\subset \!\mathbb{R}^n$, control $u: \mathbb{R}_{\geq 0} \!\to\! \mathcal{U} \!\subset\! \mathbb{R}^m$, $\mathcal{U}$ compact and known vector field $f \in C( \mathcal{X} \times \mathcal{U}; \mathbb{R}^n)$ Lipschitz ($Lip$) in the first argument. The unknown flow $r \in Lip(\mathcal{X}; \mathbb{R}^n)$ is a sample of a multi-output Gaussian process (GP)\footnote{A GP is a random process in which every finite selection of random variables is multivariate normal distributed. Mean and covariance of these random vectors are determined by a mean function $\mu$ and a positive definite kernel $k$. For an extensive introduction into GPs see~\cite{rasmussen2005gaussian}.} defined on $\mathcal{X}$: $r \sim \mathcal{GP}(\mu(x), k(x,x'))$ with $\mu \in Lip(\mathcal{X}; \mathbb{R}^n)$ and a diagonal matrix valued kernel with diagonal entries $k_j(x,x') \in C^2(\mathcal{X} \times \mathcal{X} ; \mathbb{R})$ s.t. its samples are Lipschitz. 
We assume that $r(x(t))$ can be evaluated anytime. \\
We will frame the active learning (AL) problem as an optimal control problem (OCP). The two AL methods we propose are based on OCPs whose objective is to find the trajectory, or more precisely its trace, which leads to the largest reduction of the posterior variance of the GP across the entire domain. The OCP is based on the GP mean as a surrogate for the unknown flow. In our first approach, the planning horizon is fixed. This translates to the following OCP: Problem~\eqref{problem:original_problem_fixed_end_time}:
\vspace{-0.5em}
\begin{subequations} 
\label{problem:original_problem_fixed_end_time} 
\begin{align}
\min  &_{u \in L^\infty([0,T];\mathcal{U}),} \int_\mathcal{X} \sum_{j=1}^n  k_{x, j}(\hat{x};\hat{x}) d\hat{x}  && \label{eq:objective_fixed_end_time}\\[-2pt] 
\text{s.t.}\ &\dot{x}(t) = f(x(t),u(t))+\mu(x(t)), \ x(0)=x_0,&& \label{eq:mean_dynamics_fixed_end_time}
\end{align} 
\end{subequations}
where $k_{x,j}(\hat{x};\hat{x})$ denotes the posterior variance in dimension $j$ at $\hat{x}$ after having observed the GP on the entire trace of $x$.
The second approach accounts for the uncertainty of the flow: Even though the true trajectory cannot be predicted surely, it is possible by the regularity assumptions about the GP to establish a probabilistic time dependent bound $d$ on the deviation of the true trajectory based on~\eqref{eq:ground_truth_control_system} from the nominal trajectory based on $\dot{x} = f(x,u) + \mu(x), \ x(0) = x_0$~\cite{tracking_error__Lederer, treven2023efficientexplorationcontinuoustimemodelbased}:  $\dot{d}(t) = L_k d(t) + \beta_k\sqrt{\sum_{j = 1}^nk_j(x(t),x(t))}, \ d(0) = 0, $
where $x$ denotes the nominal trajectory, $\beta_k$ and $L_k$ originate from uniform prediction error bounds on the GP and its derivative, and the sum bounds the operator norm of the kernel matrix. Notice that $d$ is a monotonously increasing.\\
Our second AL approach involves the solution an free end time OCP that only admits bounded deviations of the true trajectory from the planned trajectory: Problem~\eqref{problem:original_problem}
\begin{subequations} 
\label{problem:original_problem} 
\begin{flalign} 
\min  &_{T \in \mathbb{R}_{>0},\, u \in L^\infty([0,T];\mathcal{U}),} \int_\mathcal{X} \sum_{j=1}^n  k_{j,x}(\hat{x};\hat{x}) d\hat{x} && \label{eq:objective}\\
\text{s.t.}\ &\dot{x}(t) = f(x(t),u(t))+\mu(x(t)), \ x(0)=x_0,&& \label{eq:mean_dynamics}\\
&\dot{d}(t) = L_k d(t) + \beta_k \sqrt{\sum_{j=1}^n k_j(x(t),x(t))}, \ d(0)\!=\!0, \!\!\!\!\!\!\!\!\!\!\! && \label{eq:uncertainty_dynamics} \\[-10pt]
& d(T)\leq M. && \label{eq:uncertainty_bound}
\end{flalign} 
\end{subequations}
\section{Transformation of the Problems}
In this section, we derive tractable surrogates for problems~\eqref{problem:original_problem_fixed_end_time} and~\eqref{problem:original_problem}. Therefore, we transform and approximate the objective and eliminate~\eqref{eq:uncertainty_dynamics},~\eqref{eq:uncertainty_bound}, and the free end time in~\eqref{eq:objective} by trajectory-wise time rescaling. 

\subsection{Objective}
Without loss of generality, we consider the GPs of all dimensions $j$ separately. 
By~\cite[Thm. 3.2]{steinwart2026conditioning}, a GP conditioned on infinitely many data points is a GP, and the posterior variance at $x_i \in \mathcal{X}$ is 
\begin{equation}
\begin{aligned}
    &\!\!\!\!\!\lim_{l \to \infty} k_j(x_i, x_i) - k_j(x_i, X_l) k_j(X_l, X_l)^{-1}k_j(X_l, x_i) \\ 
    &\!\!\!\!\!= \lim_{l \to \infty}\min_{a \in \mathbb{R}^l} \|k_j(\cdot, x_i) \!- \!k_j(x_i, X_l)a\|_{\mathcal{H}_j}^2=: k_{x,j}(x_i, x_i), \label{eq:posterior_of_curve_limit}
\end{aligned}
\end{equation}
where $\mathcal{H}_j$ with norm $\| \cdot\|_{\mathcal{H}_j}$ denotes the reproducing kernel Hilbert space (RKHS) belonging to the kernel $k_j$ and $(X_l)_{l = 1}^\infty$ is an inclusion-wise monotonously growing sequence of sets with cardinality $|X_l| = l$ becoming dense in $x([0,T]) \! := \!\{\hat{x}\! \in \!\mathcal{X}| \exists t \!\in\! [0,T]\!: \!\hat{x}\! = \!x(t)\}$, i.e. the trace of the trajectory. The following subspaces of $\mathcal{H}_j$ are constructed over the trace $x([0,T])$ resulting from~\eqref{eq:mean_dynamics_fixed_end_time} and~\eqref{eq:mean_dynamics}, respectively:
    \setlength{\abovedisplayskip}{2pt}
    \setlength{\belowdisplayskip}{2pt}
\begin{align*}
    \mathcal{H}_{x,j, pre} &:= \! \left\{ \sum_{l = 1}^n \alpha_i k_j(\cdot, x_l) : n \!\in \!\mathbb{N}, \alpha_l \!\in\! \mathbb{R}, x_l \!\in\! x([0,T])\! \right\}\!, \\[-3pt]
    \mathcal{H}_{x,j} &:= \overline{\mathcal{H}_{x, j, pre}}^{\| \cdot \|_{\mathcal{H}_j}}, 
\end{align*}
where the last term denotes the closure of $\mathcal{H}_{x, pre}$ w.r.t. the norm of $\mathcal{H}_j$.
Hence,~\eqref{eq:posterior_of_curve_limit} is equal to $\inf_{g \in  \mathcal{H}_{x,j, pre}} \|f -g\|_{\mathcal{H}_j}^2$ (as $k_j \in C$) which is again equal to $\min_{g \in  \mathcal{H}_{x,j}} \|f -g\|_{\mathcal{H}_j}^2$ as $\mathcal{H}_{x,j, pre}$ is dense in $\mathcal{H}_{x,j}$. The minimization is well posed as it is a projection onto a closed Hilbert subspace. We denote this projection by $P_{\mathcal{H}_{x,j}} \!:\! \mathcal{H}_j \! \to \! \mathcal{H}_j, \ f \mapsto \arg\min_{g \in  \mathcal{H}_{x,j}} \|f -g\|_{\mathcal{H}_j}^2$.  In summary, we obtain
    \setlength{\abovedisplayskip}{2pt}
    \setlength{\belowdisplayskip}{2pt}
\begin{equation}
\refstepcounter{equation}
    \begin{aligned}
        & k_{x,j}(x_i, x_i) &= \inf_{g \in  \mathcal{H}_{x,j, pre}} \!\!\!\|k_j(\cdot, x_i) -g\|_{\mathcal{H}_j}^2 && \!\!\!\!\!\!\!\!\!\!\\[-3pt]
        \!\!\!\!= \ \! & \|k_j(\cdot, x_i) \!-\!P_{\mathcal{H}_{x,j}}k_j(\cdot, x_i)\|_{\mathcal{H}_j}^2 \!\!\!\!\!\!\!\!\!\!\!\! &= \!\min_{g \in  \mathcal{H}_{x,j}} \|k_j(\cdot, x_i) -g\|_{\mathcal{H}_j}^2. && \!\!\!\!\!\!\!\!\!
    \end{aligned}
\label{eq:equallity_projection}
\tag{\theequation a}
\end{equation}
Next, we want to extend these equalities to a continuous-time setting. Therefore, we define 
\begin{align*}
    \tilde{\mathcal{H}}_{x,j, pre} &:= \left\{ \int_0^T k_j( \cdot , x(t))  a(t) dt \bigg| a \in L^2([0,T], \mathbb{R}) \right\}, \\[-3pt]
    \tilde{\mathcal{H}}_{x,j} &:= \overline{\tilde{\mathcal{H}}_{x,j, pre}}^{\| \cdot \|_{\mathcal{H}_j}},
\end{align*}
where the integral is in the sense of Bochner and it is well defined by~\cite[Prop. 1.2.2]{Hytonen2016BanachSpacesI} as $\|\|k_j( \cdot , x(t))  a(t)\|_{\mathcal{H}_j}\|_{L^1} = \|\sqrt{k_j( x(t), x(t))}|a(t)\|_{L^1} \leq \|\sqrt{k_j(x(t), x(t))}\|_{L^2}  \| a\|_{L^2}$. 
This bound is finite by the continuity of $k_j$ and $x$.  
\begin{proposition}
\label{thm:equation_sub_rkhs}
    If $k_j$ is continuous, 
    $\mathcal{H}_{x,j} = \tilde{\mathcal{H}}_{x,j}$.\footnote{\label{proofs_in_appendix}Proof can be found in the appendix.}
\end{proposition}
Hence, we can extend~\eqref{eq:equallity_projection} by
    \setlength{\abovedisplayskip}{2pt}
    \setlength{\belowdisplayskip}{2pt}
\begin{equation}
\begin{aligned}
    k_{x,j}(x_i, x_i) &= \|k_j(\cdot, x_i) \!-\!P_{\tilde{\mathcal{H}}_{x,j}}k_j(\cdot, x_i)\|_{\mathcal{H}_j}^2\\[-2pt]
    &= \|P_{\tilde{\mathcal{H}}_{x,j}^\perp}k_j(\cdot, x_i)\|_{\mathcal{H}_j}^2,
\end{aligned}
\tag{\theequation b} \label{eq:equalities_continued}
\end{equation}
where $\tilde{\mathcal{H}}_{x,j}^\perp$ denotes the orthogonal complement of $\tilde{\mathcal{H}}_{x,j}$. \newpage
Leveraging the observation~\eqref{eq:equalities_continued}, it follows that the integral in the objective in~\eqref{eq:objective_fixed_end_time} and~\eqref{eq:objective}, respectively, can be approximated by quadrature as the integrand is continuous: 
$\sum_{j=1}^n \|P_{\tilde{\mathcal{H}}_{x,j}^\perp}k_j(\cdot, x_i)\|_{\mathcal{H}_j}^2$ is a continuous function because $\left\|P_{(\mathcal{H}_{x,j})^\perp} \bullet \right\|_{\mathcal{H}_{j}}^2$ is a composition of continuous functions and $\hat{x} \mapsto k_j(\cdot; \hat{x})$ is continuous, too, 
because $\|k_j(\cdot; \hat{x}) \! - \!k_j(\cdot; \hat{y})\|_{\mathcal{H}_j} = \sqrt{k_j(\hat{x}, \hat{x})\! + \!k_j(\hat{y}, \hat{y})\! - \!2k_j(\hat{x}, \hat{y})} \xrightarrow{\hat{x} \to \hat{y}} 0$ by continuity of $k_j$. Hence, in~\eqref{eq:objective_fixed_end_time} and~\eqref{eq:objective} we approximate
    \setlength{\abovedisplayskip}{2pt}
    \setlength{\belowdisplayskip}{2pt}
\begin{equation}
\label{eq:approximated_objective}
    \int_\mathcal{X} \sum_{j=1}^n  k_{x, j}(\hat{x};\hat{x}) d\hat{x} \approx \sum_{i = 1}^N w_i \sum_{j=1}^n  \|P_{\tilde{\mathcal{H}}_{x,j}^\perp}k_j(\cdot, x_i)\|_{\mathcal{H}_j}^2, 
\end{equation}
$w_i \! \in \! \mathbb{R}_{\geq 0}$ being quadrature weights and $x_i \! \in \! \mathcal{X}$ test points. 

The summands in~\eqref{eq:approximated_objective} are still not tractable. Because of that, we derive a tractable approximation scheme for them. Therefore, we define the operator 
    \setlength{\abovedisplayskip}{2pt}
    \setlength{\belowdisplayskip}{2pt}
\[ G_{x,j}:L_2([0,T],\mathbb{R}) \rightarrow \mathcal{H}_j, \ a\mapsto \int_0^T k_j(\cdot,x(t))a(t)\,dt, \]
where the integral is understood in the sense of Bochner. We assert the following two properties of $G_{ij}$:
\begin{proposition} 
    \label{thm:adjoint_operator}
    The adjoint operator of \(G_{x,j}\) is given by\footref{proofs_in_appendix}  \[ G_{x,j}^*:\mathcal{H}_j\rightarrow L_2([0,T]), \qquad (G^*f)(t)=f(x(t)).\]
\end{proposition} 

\begin{proposition}
\label{thm:G_compact}
    Assume that the kernel \(k_j\) is continuous. Then the operators $G_{x,j}^*, G_{x,j}$ and $G_{x,j}^*G_{x,j}$ are compact.\footref{proofs_in_appendix} 
\end{proposition}
Notice that 
$\argmin_{a \in  L^2([0,T])} \|f -G_{x,j}a\|_\mathcal{H}$ is not well defined
as $G_{x,j}$ is a compact operator with infinite dimensional range for some kernels e.g. the RBF kernel.
Recall that this problem is well-posed by Lax-Milgram
\begin{equation}
\label{eq:tikhunov}
    \min_{a_\alpha\in L_2} \|k_j(\cdot;x_i)-G_{x,j}a_\alpha\|_{\mathcal{H}_j}^2 + \alpha\|a_\alpha\|_{L_2}^2,
\end{equation}
and that the unique minimizer is characterized by
\begin{equation}
\label{eq:variational_equation}
\begin{aligned}
     G_{x,j}^*k_j(\cdot,x_i)&= (G_{x,j}^*G_{x,j}+\alpha I)a_\alpha.
     \\
     \text{i.e. } k_j(x(\cdot), x_i)&= \int_0^T k_j(x(\cdot), x(t))a_\alpha(t)dt + \alpha a_\alpha(\cdot).
\end{aligned}
\end{equation}
The operator on the right is a positive definite compact perturbation of the identity and hence invertible~\cite[Thm. 3.4]{Kress2014}. We denote the solution of~\eqref{eq:variational_equation} by $a_\alpha =: a_{x, j, \alpha}(\cdot; x_i)$. 
By~\cite[Thm. 15.23., Thm. 16.5.]{Kress2014} it holds:
\begin{equation}
    G_{x,j}a_{x, j, \alpha}(\cdot; x_i) \rightarrow P_{\overline{\operatorname{ran}(G_{x,j})}}k_j(\cdot, x_i), \quad \text{for } \alpha\to 0. 
\end{equation}
So \eqref{eq:tikhunov} approximates $\|P_{(\mathcal{H}_{x,j})^\perp} k_j(\cdot, x_i)\|_{\mathcal{H}_j}^2$ for $\alpha \to 0$ and it can be rewritten to
\begin{align}
    & \!\left\|  k_j(\cdot,x_i)\! -\!\! \int_0^T \!\!\!\!k_j(\cdot,x(t))a_{x,j, \alpha}(t;x_i) dt \right\|_{\mathcal{H}_{k_j}}^{2} \! \!\!\!\!\!\!\!\!\! + \!\alpha \! \left\| a_{x,j, \alpha}(\cdot;x_i) \right\|_{L^2}^2\notag \\[-4pt]
    &= k_j(x_i , x_i) - 2 \int_0^Tk_j(x_i, x(t))a_{x,j,\alpha}(t;x_i) dt \notag \\[-4pt]
    & + \int_0^T \int_0^T a_{x,j,\alpha}(s;x_i)k_j(x(s),x(t)) a_{x,j,\alpha}(t;x_i) dsdt
    \notag \\[-4pt]
    &\overset{\eqref{eq:variational_equation}}{=} k_j(x_i , x_i) -  \int_0^Tk_j(x_i, x(t))a_{x,j,\alpha}(t;x_i)dt. \label{eq:objectiv_2}
\end{align}

Leveraging these findings, we rewrite~\eqref{problem:original_problem_fixed_end_time} by Problem~\eqref{problem:fixed_endtime_2}:
    \setlength{\abovedisplayskip}{2pt}
    \setlength{\belowdisplayskip}{2pt}
\begin{subequations}
    \label{problem:fixed_endtime_2}
    \begin{align}
    \min  &_{u \in L^\infty([0,T]; \mathcal{U})}\notag\\
    \lim_{\alpha \to 0} &
        \sum_{i=1}^N \! w_i \! \sum_{j = 1}^n k_j(x_i , x_i)  - \! \! \int_0^M \!\!\!\! k_j(x_i, x(t))a_{x, j,\alpha}(t;x_i)dt  \label{eq:fixed_endtime_2_objective} \\[-10pt]
        s.t.  & \int_0^T k(x(t), x(s)) a_{x, j,\alpha}(s ; x_i)ds + \alpha a_{x, j,\alpha}(t; x_i) \notag \\
        &= k(x(t), x_i), \ a.e. \ t \in [0,T], \ \forall i \in [n]\label{eq:fixed_endtime_2_variational_equation} \\
    \dot{x}(t) &= f(x(t), u(t)) + \mu(x(t)), x(0) = x_0. \!\!\! \label{eq:fixed_endtime_2_mean_dynamics}
\end{align}
\end{subequations}
Compared to Problem~\ref{problem:original_problem_fixed_end_time}, the objective~\eqref{eq:fixed_endtime_2_objective} is computable and the variational equation solvable by numerical methods~\cite{Kress2014}. Furthermore, Problem~\ref{problem:fixed_endtime_2} allows to calculate gradients w.r.t. to the control as we will see in Section~\ref{sec:gradients}.

\subsection{Dynamics}
Next, we transform Problem~\eqref{problem:original_problem} into a fixed end time problem through trajectory-wise time rescaling w.r.t. $d$. For that, observe that the objective value in~\eqref{eq:objective} only depends on the trace of the trajectory but not on its temporal development.
Hence, the objective value does not change if we traverse the trajectory at a different speed.
\begin{theorem} \label{thm:time_rescaling} Let $f \in C(\mathcal{X}\times\mathcal{U};\mathbb{R}^n)$ and Lipschitz in the first argument, $\mu \in Lip(\mathcal{X};\mathbb{R}^n)$ and $\theta \in C^1(\mathcal{X};\mathbb{R}_{\geq 0})$ such that $\theta(x_0) > 0$. \(u \in L^\infty([0,T];\mathcal{U})\) is a given control function. It holds that \(x:[0,M]\to\mathcal{X}\) and \(d:[0,T]\to[0,M]\) solve
    \setlength{\abovedisplayskip}{2pt}
    \setlength{\belowdisplayskip}{2pt}
\begin{align} 
x'(s) &= \frac{ f\bigl(x(s),u(d^{-1}(s))\bigr) + \mu\bigl(x(s)\bigr) }{ \theta\bigl(x(s)\bigr)+Ls }, \ x(0)=x_0, \label{eq:rescaled_gamma} \\
d'(t) &= \theta\bigl(x(d(t))\bigr)+Ld(t), \ d(0)=0, \ d(T)=M. \label{eq:time_rescaling} \end{align} 
if and only if $x(d(\cdot))$ and $d(\cdot)$ solve
\begin{align} 
x'(t) &= f\bigl(x(t),u(t)\bigr)+\mu\bigl(x(t)\bigr), \ x(0)=x_0, \label{eq:original_dynamics} \\ 
d'(t) &= \theta\bigl(x(t)\bigr)+Ld(t), \ d(0)=0, \ d(T)=M, \label{eq:original_time_rescaling} 
\end{align}
for all \(t\in[0,T]\).
\end{theorem}
This gives rise to a reformulation of~\eqref{problem:original_problem}:
    \setlength{\abovedisplayskip}{2pt}
    \setlength{\belowdisplayskip}{2pt}
\begin{subequations}
\label{problem:rescaled_problem}
    \begin{align}
    \min  &_{u \in L^\infty([0,M]; \mathcal{U})}\notag\\[-3pt]
    \lim_{\alpha \to 0} &
        \sum_{i=1}^N \! w_i \! \sum_{j = 1}^n k_j(x_i , x_i)  - \! \! \int_0^M \!\!\!\! k_j(x_i, x(t))a_{x, j,\alpha}(t;x_i)dt  \label{eq:rescaled_objective} \\[-8pt]
        s.t.  & \int_0^M k(x(t), x(s)) a_{x, j,\alpha}(s ; x_i)ds + \alpha a_{x, j,\alpha}(t; x_i) \notag \\[-3pt]
        &= k(x(t), x_i), \ a.e. \ t \in [0,T], \ \forall i \in [n]\label{eq:rescaled_variational_equation} \\
    \dot{x}(t) &= \frac{f(x(t), u(t)) + \mu(x(t))}{L_kt + \beta_k \sqrt{\sum_{j = 1}^n k_j(x(t), x(t)) }}, x(0) = x_0. \!\!\! \label{eq:rescaled_mean_dynamics}
\end{align}
\end{subequations}
Indeed, this is equivalent to~\eqref{problem:original_problem}:
\begin{theorem}
\label{thm:rescaling_equivalence}
    Problem~\eqref{problem:original_problem} and problem~\eqref{problem:rescaled_problem} are equivalent, i.e. they have the same minimal objective value, the same optimizing trajectory and if $u^*(\cdot)$ is the optimal control in~\eqref{problem:original_problem} then $u^*(d^{-1}(\cdot))$ is the OC for~\eqref{problem:rescaled_problem} with $d$ from~\eqref{eq:original_time_rescaling} and if $u^*(\cdot)$ is the optimal control in~\eqref{problem:rescaled_problem} then $u^*(d(\cdot))$ is the OC for~\eqref{problem:original_problem} with $d$ from~\eqref{eq:time_rescaling}.
\end{theorem}
\begin{proof}
    The dynamics parts in~\eqref{problem:original_problem} and~\eqref{problem:rescaled_problem} can produce exactly the same set of traces of trajectories by~\eqref{thm:time_rescaling}. 
    As the objective value depends only on the trace of the trajectories, also the minimal possible objective value in~\eqref{problem:original_problem} and~\eqref{problem:rescaled_problem} are the same and connected to the same trajectory trace. Further, if an optimal trajectory in~\eqref{problem:rescaled_problem} is a roll-out of a control $u: [0,M] \to \mathcal{U}$, then $u(d(\cdot))$  with $d$ from~\eqref{eq:time_rescaling} leads to the same trajectory trace based on~\eqref{eq:mean_dynamics} by~\ref{thm:time_rescaling}. The other implication in dealt with similarly. 
\end{proof}
By means of the above theorem, one can solve the simpler, fixed end time OCP~\eqref{problem:rescaled_problem} with fewer side conditions instead of~\eqref{problem:original_problem}. Therefore, one computes the optimal control $u^*$ and trajectory $x^*$ from~\eqref{problem:rescaled_problem} at first and solves then~\eqref{eq:time_rescaling} for that $x^*$. The solution of~\eqref{problem:original_problem} is then $u^*(d(\cdot))$.

\section{Gradients \& Pontryagin's Maximum Principle}
\label{sec:gradients}
In this section, we show for problems~\eqref{problem:fixed_endtime_2} and~\eqref{problem:rescaled_problem} with $\alpha \! >\! 0$ fixed, that the objective value is differentiable w.r.t. the control and how the gradients are calculated. Based on that, we derive a Pontryagin-type necessary condition for the optimal control. For notational simplicity, we denote the vector field in~\eqref{eq:fixed_endtime_2_mean_dynamics} and~\eqref{eq:rescaled_mean_dynamics} by $\tilde{f}(t, x,u)$.
We show this differentiability chain: Objectives~\eqref{eq:fixed_endtime_2_objective} and~\eqref{eq:rescaled_objective} are Fréchet differentiable w.r.t. the trajectory $x$ and the regressors $a_{x, j, \alpha}$ (Thm.~\ref{thm:derivative_objective}), $a_{x, j, \alpha}$ is differentiable w.r.t. $x$ by means of~\eqref{eq:fixed_endtime_2_variational_equation} and~\eqref{eq:rescaled_variational_equation} (Thm.~\ref{thm:differentiability_of_variational_equation}). The differentiability of $x$ w.r.t. $u$ under~\eqref{eq:fixed_endtime_2_mean_dynamics} and~\eqref{eq:rescaled_mean_dynamics} is well known~\cite[Thm. 6.1.1]{bressan2007introduction}.
\begin{theorem}
    \label{thm:derivative_objective}
    Take the kernels $k_j \in C^2$, the trajectory $x$ continuous and $a_{x,j,\alpha} \in L^2$. Then the Fréchet derivative of the objective~\eqref{eq:rescaled_objective} is\footref{proofs_in_appendix} 
        \setlength{\abovedisplayskip}{2pt}
    \setlength{\belowdisplayskip}{2pt}
    \begin{equation*}
        -\! \sum_{i=1}^N \! w_i \! \sum_{j = 1}^n \!  \! \int_0^M \! \! \! \! \!\! \!  \partial_2k_j(x_i, x(t))h(t)a_{x,j,\alpha}(t)\! 
        +\!   k_j(x_i, x(t))\xi(t) dt.
    \end{equation*}
\end{theorem}

Next we differentiate $a_{x,j,\alpha}$ w.r.t. $x$ for an $\alpha > 0$ by means of~\eqref{eq:rescaled_variational_equation}.
Therefore, we define
    \setlength{\abovedisplayskip}{2pt}
    \setlength{\belowdisplayskip}{2pt}
\begin{equation*}
\begin{aligned}
    g_{ij}:  & \ C([0,M];\mathbb R^n) \times L_2([0,M]) \rightarrow L_2([0,M]), \\ 
    & \! \!  (x,a) \!  \mapsto \!  \!  \int_0^M \! \! \! k_j(x(\cdot),x(s)) a(s)\,ds + \alpha a(\cdot) - k_j(x(\cdot),x_i),
\end{aligned}
\end{equation*}
and observe that~\eqref{eq:variational_equation} is equivalent to $g_{ij}(x,a_{x,j,\alpha}(\cdot, x_i)) \overset{a.e.}{=}  0$. This yields the required differentiability: 
\begin{theorem}
\label{thm:differentiability_of_variational_equation}
Let $(x,a)$ satisfy $g_{ij}(x,a)=0$
and $k_j \in C^2(\mathcal{X} \times \mathcal{X}; \mathbb{R})$ positive definite. Then there exist open neighbourhoods $x \in U\subset C([0,M];\mathbb R^n), \ a \in V\subset L_2([0,M])$ and a Fréchet differentiable map $l_{ij}:U\rightarrow V,$ such that $g_{ij}(x',l_{ij}(x'))=0, \forall x'\in U.$ Moreover, $g_{ij}(x',a')=0 \Longleftrightarrow a'=l_{ij}(x') \quad \forall  (x',a')\in U\times V.$\footref{proofs_in_appendix} 
\end{theorem}

Having the derivatives, we can derive a first-order necessary condition for the optimal solution of~\eqref{problem:fixed_endtime_2} and~\eqref{problem:rescaled_problem}:
\begin{theorem}[Maximum Principle] \label{thm:PMP}
Define \(v\) by
    \setlength{\abovedisplayskip}{2pt}
    \setlength{\belowdisplayskip}{2pt}
\begin{equation}
     \label{eq:variational_dynamics_statement}
     \begin{aligned}
         \dot v(t) &= \tilde{f}_x(t, x^*(t),u^*(t))v(t),\\  
     v(\hat t)& = \tilde{f}(\hat{t}, x^*(\hat t),\omega) - \tilde{f}(\hat t, x^*(\hat t),u^*(\hat t)). 
     \end{aligned}
\end{equation}
for $t \in [\hat{t}, M]$ and $v(t) = 0$ for $t \in [0,\hat{t})$. 
Let \((u^*,x^*,(a_{ij}^*)_{i,j}^{N,n})\) be an optimal solution of~\eqref{problem:fixed_endtime_2}, ~\eqref{problem:rescaled_problem}, respectively, with a fixed $\alpha >0$. If \(q:[0,M]\to\mathbb R^n\) satisfies
    \setlength{\abovedisplayskip}{2pt}
    \setlength{\belowdisplayskip}{2pt}
\begin{align}
     -\dot{q}(t)^{T}v(t) = \ &  q(t)^{T} \tilde{f}_x\bigl(t,x^*(t),u^*(t)\bigr)v(t) \label{eq:adjoint_equation}\\[-3pt]
     &  + \sum_{i=1}^{N}w_i \sum_{j=1}^{m} \partial_2 k_j\bigl(x_i,x^*(t)\bigr)v(t)\,a_{ij}^*(t) \notag \\[-3pt]
     & + k_j\bigl(x_i,x^*(t)\bigr) D_{x^*,a_{ij}^*}l_{ij}(v)(t) \notag
\end{align}
and $q(M) = 0$ for all $\omega \in \mathcal{U}$ and $\hat{t} \in [0,M]$, then for a.e. \(\hat t\in[0,M]\) and every \(\omega\in \mathcal{U}\),\footref{proofs_in_appendix} 
    \setlength{\abovedisplayskip}{2pt}
    \setlength{\belowdisplayskip}{3pt}
\[ q(\hat t)^T \tilde{f}(\hat{t}, x^*(\hat t),\omega) \le q(\hat t)^T \tilde{f}(\hat{t}, x^*(\hat t),u^*(\hat t )). \]
\end{theorem}
The results of this section allow us to solve Problems~\ref{problem:fixed_endtime_2} and~\ref{problem:rescaled_problem} with gradient-based optimization techniques and to check that the found solution is indeed locally optimal.
\section{Experiments}
In this section, we compare our methods against \textit{rec} and \textit{p\&a} from~\cite{pmlr-v120-buisson-fenet20a} on the basis of three numerical examples\footnote{Code basis: \url{https://gitlab.lrz.de/000000000149EE60/active_learning_of_flows_from_curve_observations_using_gaussian_processes.git}}. Thereby, the two benchmarks \textit{rec} and \textit{p\&a} are varied such that the domain of the GPs is just $\mathcal{X}$ and not $\mathcal{X} \times \mathcal{U}$.
\subsection{Setup}
In both experiments, the flow $r$ of the spring-mass-damper system
$\dot{x} \! =\! f(x, u) \!+ \!r(x), \!\ x_0\! = \!(1,0)^T\!, \!\ u \!\in \![-5, 5],\!$
with
$f(x_1, x_2,u) \! = \!(0,u)^T, \ r(x_1,x_2) \!=\! (x_2, - 2 x_1 - 0.01x_2
    )^T\!,$
has to be identified on the domain $\mathcal{X}' := [-10, 10]^2 \subset \mathcal{X}$. In the first experiment, the algorithms only know $f$ initially; in the second, additionally, 100 noise-free measurements of $r$ on $[-3, 3]^2$ are given in advance; and
in the third, 120 noise-free measurements of $r$ on $\mathcal{I}\!\! := \!\!\{4 \!\leq \!\sqrt{x_1^2\! + \!x_2^2} \!\leq 6\! \}$ are given in advance.
$\textit{rec}$ and $p\&a$~\cite{pmlr-v120-buisson-fenet20a} solve
    \setlength{\abovedisplayskip}{3pt}
    \setlength{\belowdisplayskip}{3pt}
\begin{equation}
\label{problem:variance_objective}
\begin{aligned}
    \!\!\!\! \argmax_{u \in \mathcal{U}^h} & \sum_{i = 1}^{h}\sum_{j = 1}^n k_{j}(x_i, x_i) \\[-3pt]
    \!\!\!\! s.t. \ & x_{i +1} = x_i + \Delta t (f(x_i, u_i) + \mu(x_i)), \ x_0 = x_{t},
\end{aligned} \!\!\!\!
\end{equation}
where $\mu_j$ and $k_j$ are the (posterior) mean and kernel of the assumed GPs that should model the components of $r$, 
and apply the solution in both an \textit{rec} (receding horizon) and \textit{p\&a} (plan and apply) fashion. While \textit{rec} solves the optimization problem in every time step and applies only the first control value, \textit{p\&a} executes the entire control sequence before it solves the optimization problem again. $h$ is the prediction horizon, $n$ the state dimension, i.e. $n = 2$. \\
Against that, we compare our four algorithms: 
Instead of~\eqref{problem:variance_objective}, our first and second methods solve Problem~\ref{problem:fixed_endtime_2} and apply its solution in a $\textit{rec}$ and in a $\textit{p\&a}$ fashion, respectively. 
Our third and forth method solve Problem~\ref{problem:rescaled_problem} with $M \! = \!2, \ L_k \!= \!2, \ \beta_k \!=\! 0.5$ and apply its solution time-rescaled by means of Theorem~\ref{thm:rescaling_equivalence} in a $\textit{rec}$ and in a $\textit{p\&a}$ fashion, respectively.
All algorithms are tested on the same 15 parameter configurations of the RBF kernels $k_{j \in \{1,2\}}$ for the GPs ($\{\sigma_1 = 10\} \times \{\sigma_2 = 5,10,20\} \times \{l_1 = l_2 = 1,2,3,4,5\}$) and always use a prediction horizon of 20 steps, Runge-Kutta 4 integration of the dynamics, and a projected gradient optimizer for their OCP. The evaluation is based on the root mean squared error (RMSE) between $r$ and the algorithm's estimate of it $\hat{r}$, i.e., $\|r -\hat{r}\|_{L^2(\mathcal{X}')}$. The total variation of $r$ is $\|r -0\|_{L^2(\mathcal{X}')} = 13.4$.

\subsection{Results}

\def\prefix{empty/RMSE_overview_controllers_6_T_40_horizon_20_time_rescaling_beta_0_5_L_2_M_2_all_experiments_}
\pgfplotstableread{\prefix 1.txt}\data
\pgfplotstablevertcat{\data}{\prefix 2.txt}
\pgfplotstablevertcat{\data}{\prefix 3.txt}
\pgfplotstablevertcat{\data}{\prefix 4.txt}
\pgfplotstablevertcat{\data}{\prefix 5.txt}

\def\firstcol{t_0}
\pgfplotstablegetcolsof{\data}
\pgfmathtruncatemacro\lastidx{\pgfplotsretval-1}
\pgfplotstablegetcolumnnamebyindex{\lastidx}\of{\data}\to\lastcol
\pgfplotstablecreatecol[create col/expr={\thisrow{\firstcol}-\thisrow{\lastcol}}]{delta}\data

\pgfplotstablesort[sort key=delta]\sorted\data
\makeatletter
\pgfplotstableforeachcolumnelement{controller}\of\sorted\as\ctrl{%
  \ifcsname n@\ctrl\endcsname\else\expandafter\xdef\csname n@\ctrl\endcsname{0}\fi
  \pgfplotstablegetelem{\pgfplotstablerow}{delta}\of\sorted
  \expandafter\xdef\csname v@\ctrl @\csname n@\ctrl\endcsname\endcsname{\pgfplotsretval}%
  \expandafter\xdef\csname n@\ctrl\endcsname{\the\numexpr\csname n@\ctrl\endcsname+1\relax}%
}

\newcommand\q[2]{%
  \ifcsname n@#1\endcsname
    \pgfmathsetmacro\qpos{(\csname n@#1\endcsname-1)*#2}%
    \pgfmathtruncatemacro\qlo{floor(\qpos)}%
    \pgfmathtruncatemacro\qhi{min(\qlo+1,\csname n@#1\endcsname-1)}%
    \pgfmathsetmacro\qval{\csname v@#1@\qlo\endcsname
      + (\qpos-\qlo)*(\csname v@#1@\qhi\endcsname-\csname v@#1@\qlo\endcsname)}%
    \pgfmathprintnumber[fixed,precision=2,zerofill]{\qval}%
  \else --\fi}
\makeatother

\newcommand\qrow[2]{#1
  & \q{GPVarianceControllerREC}{#2}                & \q{GPVarianceControllerPAA}{#2}
  & \q{GPInformationControllerFixedEndTimeREC}{#2} & \q{GPInformationControllerFixedEndTimePAA}{#2}
  & \q{GPInformationControllerFreeEndTimeREC}{#2}  & \q{GPInformationControllerFreeEndTimePAA}{#2} \\}

\begin{table}[htbp]
\centering
\begin{tabular}{lcccccc}
  \toprule
  & \multicolumn{2}{c}{OCP~\eqref{problem:variance_objective}}
  & \multicolumn{2}{c}{OCP~\eqref{problem:fixed_endtime_2}}
  & \multicolumn{2}{c}{OCP~\eqref{problem:rescaled_problem}} \\
  \cmidrule(lr){2-3} \cmidrule(lr){4-5} \cmidrule(lr){6-7}
  & rec & p\&a & rec & p\&a & rec & p\&a \\
  \midrule
  \qrow{best}{1}
  \qrow{1st quartile}{0.75}
  \qrow{median}{0.5}
  \qrow{3rd quartile}{0.25}
  \qrow{worst}{0}
  \bottomrule
\end{tabular}
\caption{Reduction of RMSE: $\mathrm{RMSE}(t_0) \!- \!\mathrm{RMSE}(t_{\mathrm{end}})$, \\ experiment without initial measurements.}
\label{tab:rmse_delta}
\end{table}
\vspace{-1em}
In the first experiment without initially given data points, the algorithms based on~\eqref{problem:variance_objective} and~\eqref{problem:fixed_endtime_2} perform similarly, whereas the methods based on~\eqref{problem:rescaled_problem} reduce the RMSE less. In general, it can be said that the \textit{rec} strategy performs better than \textit{p\&a}. This confirms with~\cite{pmlr-v120-buisson-fenet20a}. 

\def\prefix{quad/RMSE_overview_controllers_6_T_40_horizon_20_time_rescaling_beta_0_5_L_2_M_2_all_experiments_}
\pgfplotstableread{\prefix 1.txt}\dataquad
\pgfplotstablevertcat{\dataquad}{\prefix 2.txt}
\pgfplotstablevertcat{\dataquad}{\prefix 3.txt}
\pgfplotstablevertcat{\dataquad}{\prefix 4.txt}
\pgfplotstablevertcat{\dataquad}{\prefix 5.txt}

\def\firstcol{t_0}
\pgfplotstablegetcolsof{\dataquad}
\pgfmathtruncatemacro\lastidx{\pgfplotsretval-1}
\pgfplotstablegetcolumnnamebyindex{\lastidx}\of{\dataquad}\to\lastcol
\pgfplotstablecreatecol[create col/expr={\thisrow{\firstcol}-\thisrow{\lastcol}}]{delta}\dataquad

\pgfplotstablesort[sort key=delta]\sortedquad\dataquad
\makeatletter
\pgfplotstableforeachcolumnelement{controller}\of\sortedquad\as\ctrl{%
  \ifcsname n@quad@\ctrl\endcsname\else\expandafter\xdef\csname n@quad@\ctrl\endcsname{0}\fi
  \pgfplotstablegetelem{\pgfplotstablerow}{delta}\of\sortedquad
  \expandafter\xdef\csname v@quad@\ctrl @\csname n@quad@\ctrl\endcsname\endcsname{\pgfplotsretval}%
  \expandafter\xdef\csname n@quad@\ctrl\endcsname{\the\numexpr\csname n@quad@\ctrl\endcsname+1\relax}%
}

\newcommand\qquadval[2]{%
  \ifcsname n@quad@#1\endcsname
    \pgfmathsetmacro\qpos{(\csname n@quad@#1\endcsname-1)*#2}%
    \pgfmathtruncatemacro\qlo{floor(\qpos)}%
    \pgfmathtruncatemacro\qhi{min(\qlo+1,\csname n@quad@#1\endcsname-1)}%
    \pgfmathsetmacro\qval{\csname v@quad@#1@\qlo\endcsname
      + (\qpos-\qlo)*(\csname v@quad@#1@\qhi\endcsname-\csname v@quad@#1@\qlo\endcsname)}%
    \pgfmathprintnumber[fixed,precision=2,zerofill]{\qval}%
  \else --\fi}
\makeatother

\newcommand\qrowquad[2]{#1
  & \qquadval{GPVarianceControllerREC}{#2}                & \qquadval{GPVarianceControllerPAA}{#2}
  & \qquadval{GPInformationControllerFixedEndTimeREC}{#2} & \qquadval{GPInformationControllerFixedEndTimePAA}{#2}
  & \qquadval{GPInformationControllerFreeEndTimeREC}{#2}  & \qquadval{GPInformationControllerFreeEndTimePAA}{#2} \\}

\begin{table}[htbp]
\centering
\begin{tabular}{lcccccc}
  \toprule
  & \multicolumn{2}{c}{OCP~\eqref{problem:variance_objective}}
  & \multicolumn{2}{c}{OCP~\eqref{problem:fixed_endtime_2}}
  & \multicolumn{2}{c}{OCP~\eqref{problem:rescaled_problem}} \\
  \cmidrule(lr){2-3} \cmidrule(lr){4-5} \cmidrule(lr){6-7}
  & rec & p\&a & rec & p\&a & rec & p\&a \\
  \midrule
  \qrowquad{best}{1}
  \qrowquad{1st quartile}{0.75}
  \qrowquad{median}{0.5}
  \qrowquad{3rd quartile}{0.25}
  \qrowquad{worst}{0}
  \bottomrule
\end{tabular}
\caption{Reduction of RMSE: $\mathrm{RMSE}(t_0) \!- \!\mathrm{RMSE}(t_{\mathrm{end}})$,\\ experiment with 100 initial measurements on $[-3,3]^2$.}
\label{tab:rmse_delta_quad}
\end{table}
\vspace{-1em}
In contrast, our proposed algorithms outperform the benchmarks in the second experiment. They are able to identify $r$ outside of $[-3, 3]^2 \subset \mathcal{X}$, too.
For most parameter configurations, the benchmarks do not improve the RMSE at all, whereas the time rescaling approach~\eqref{problem:rescaled_problem} decreases the RMSE even for worse parameter choices. \\
\textit{Remark: In the second and third experiment $RMSE(t_0)$ includes already the measurements given beforehand.} 

\def\prefix{circle/RMSE_overview_controllers_6_T_40_horizon_20_time_rescaling_beta_0_5_L_2_M_2_all_experiments_}
\pgfplotstableread{\prefix 1.txt}\datacirc
\pgfplotstablevertcat{\datacirc}{\prefix 2.txt}
\pgfplotstablevertcat{\datacirc}{\prefix 3.txt}
\pgfplotstablevertcat{\datacirc}{\prefix 4.txt}
\pgfplotstablevertcat{\datacirc}{\prefix 5.txt}

\def\firstcol{t_0}
\pgfplotstablegetcolsof{\datacirc}
\pgfmathtruncatemacro\lastidx{\pgfplotsretval-1}
\pgfplotstablegetcolumnnamebyindex{\lastidx}\of{\datacirc}\to\lastcol
\pgfplotstablecreatecol[create col/expr={\thisrow{\firstcol}-\thisrow{\lastcol}}]{delta}\datacirc

\pgfplotstablesort[sort key=delta]\sortedcirc\datacirc
\makeatletter
\pgfplotstableforeachcolumnelement{controller}\of\sortedcirc\as\ctrl{%
  \ifcsname n@circ@\ctrl\endcsname\else\expandafter\xdef\csname n@circ@\ctrl\endcsname{0}\fi
  \pgfplotstablegetelem{\pgfplotstablerow}{delta}\of\sortedcirc
  \expandafter\xdef\csname v@circ@\ctrl @\csname n@circ@\ctrl\endcsname\endcsname{\pgfplotsretval}%
  \expandafter\xdef\csname n@circ@\ctrl\endcsname{\the\numexpr\csname n@circ@\ctrl\endcsname+1\relax}%
}

\newcommand\qcircval[2]{%
  \ifcsname n@circ@#1\endcsname
    \pgfmathsetmacro\qpos{(\csname n@circ@#1\endcsname-1)*#2}%
    \pgfmathtruncatemacro\qlo{floor(\qpos)}%
    \pgfmathtruncatemacro\qhi{min(\qlo+1,\csname n@circ@#1\endcsname-1)}%
    \pgfmathsetmacro\qval{\csname v@circ@#1@\qlo\endcsname
      + (\qpos-\qlo)*(\csname v@circ@#1@\qhi\endcsname-\csname v@circ@#1@\qlo\endcsname)}%
    \pgfmathprintnumber[fixed,precision=2,zerofill]{\qval}%
  \else --\fi}
\makeatother

\newcommand\qrowcirc[2]{#1
  & \qcircval{GPVarianceControllerREC}{#2}                & \qcircval{GPVarianceControllerPAA}{#2}
  & \qcircval{GPInformationControllerFixedEndTimeREC}{#2} & \qcircval{GPInformationControllerFixedEndTimePAA}{#2}
  & \qcircval{GPInformationControllerFreeEndTimeREC}{#2}  & \qcircval{GPInformationControllerFreeEndTimePAA}{#2} \\}

\begin{table}[htbp]
\centering
\begin{tabular}{lcccccc}
  \toprule
  & \multicolumn{2}{c}{OCP~\eqref{problem:variance_objective}}
  & \multicolumn{2}{c}{OCP~\eqref{problem:fixed_endtime_2}}
  & \multicolumn{2}{c}{OCP~\eqref{problem:rescaled_problem}} \\
  \cmidrule(lr){2-3} \cmidrule(lr){4-5} \cmidrule(lr){6-7}
  & rec & p\&a & rec & p\&a & rec & p\&a \\
  \midrule
  \qrowcirc{best}{1}
  \qrowcirc{1st quartile}{0.75}
  \qrowcirc{median}{0.5}
  \qrowcirc{3rd quartile}{0.25}
  \qrowcirc{worst}{0}
  \bottomrule
\end{tabular}
\caption{Reduction of RMSE: $\mathrm{RMSE}(t_0) \!- \!\mathrm{RMSE}(t_{\mathrm{end}})$, \\ experiment with 120 measurements points on $\mathcal{I}$.}
\label{tab:rmse_delta_circle}
\end{table}
In the third experiment, where the state has to cross $\mathcal{I}$ to identify the flow properly, the algorithm based on~\eqref{problem:fixed_endtime_2} outperforms the others in the better quartiles, and the algorithms based on~\eqref{problem:rescaled_problem} outperform the others in the worse quartiles. 
  \begin{figure}[htbp]
  \centering
  \subfloat[OCP~\eqref{problem:variance_objective}, \textit{rec}, $t=9.1$\label{fig:P1}]{%
    \includegraphics[width=0.23\textwidth]{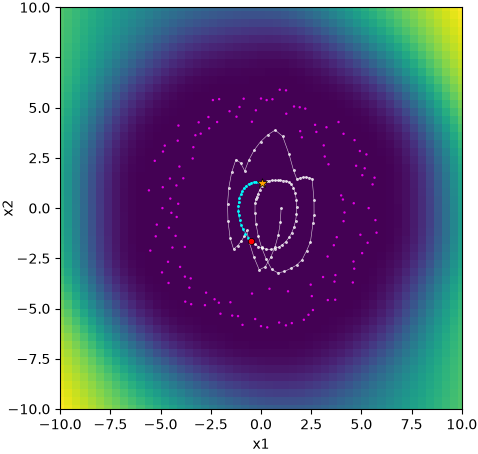}}
  \hfil
  \subfloat[OCP~\eqref{problem:rescaled_problem}, \textit{rec}, $t=1.5$\label{fig:P3}]{%
    \includegraphics[width=0.23\textwidth]{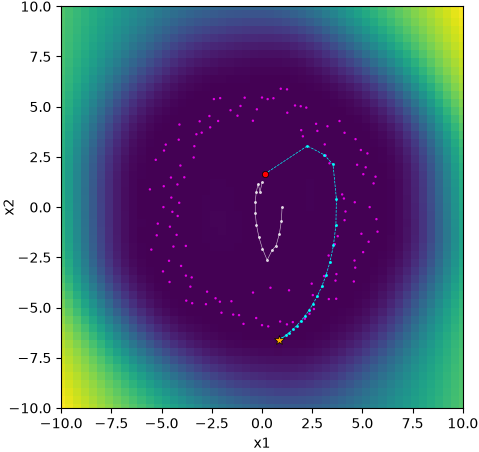}}
  \end{figure}
The figures~\ref{fig:P1} and~\ref{fig:P3} illustrate the results 
from the third experiment where the 120 data points (\symKnown) are known in advance. The two plots show where the state currently is (\symCurrent), the passed states (\symPast), the planned states (\symPlanned), and 
the target (\symTarget). 
The heatmap shows the current $l_2$ error between truth and estimate, blue meaning $0$ error, yellow $25$. The planner that uses OCP~\eqref{problem:variance_objective} does not plan trajectories crossing $\mathcal{I}$, whereas the planner using OCP~\eqref{problem:rescaled_problem} plans trajectories crossing~\eqref{problem:rescaled_problem} already after a short time. Therefore, the planned path goes along the very well-identified $\mathcal{I}$, allowing the optimizer to plan a longer path. This confirms our claims that inherently adaptive path lengths and the information-based objective~\eqref{eq:rescaled_objective} are beneficial, as unidentified points across the entire domain induce gradients, whereas the objective in~\eqref{problem:variance_objective} is only local. \\
Both proposed methods escape regions of locally maximal variance as well as well-identified regions where the variance gradient vanishes, allowing them to identify regions farther away. Which of the two is preferable depends on the scenario: if informative regions lie beyond the reach of a fixed horizon, the adaptive path length of OCP~\eqref{problem:rescaled_problem} pays off; otherwise, the fixed-horizon OCP~\eqref{problem:fixed_endtime_2} explores faster.

\section{Conclusion}
In active learning for flow identification, information-based trajectory planning criteria are able to perform better than variance-based criteria - in particular when the state is in an unidentified region. The length of the planned trajectory is crucial for prospective error reduction. In practice, trajectory-wise time rescaling leads to a genuinely adaptive method that obtains informative data while staying in already identified regions, whereas fixed end time methods explore more aggressively and hence faster.

\bibliographystyle{IEEEtran}
\bibliography{references_save}

\section*{Use of Artificial Intelligence}
We used Microsoft 365 Copilot with auto selected LLM to prove Theorems~\ref{thm:adjoint_operator},~\ref{thm:G_compact} and~\ref{thm:derivative_objective}. For Theorem~\ref{thm:derivative_objective}, we provided step 1 of our proof of~\ref{thm:differentiability_of_variational_equation} as context. During the preparation of this paper, we used Microsoft 365 Copilot and Claude Opus 5.5 and Sonnet 5.5, DeepL and Grammarly for support in structuring literature reviews, language editing and generating initial summaries of individual papers and coding.
\appendix
\renewcommand{\thesection}{A}
\refstepcounter{section}
\begin{proof}(Prop. \ref{thm:equation_sub_rkhs})
    Firstly, we show that $\tilde{\mathcal{H}}_{x,j, pre} \subset \mathcal{H}_{x,j}$.
    Therefore, we show that for a fixed $a \in L^2$, $\int_0^T k_j( \cdot , x(t))  a(t) dt$ can be approximated arbitrarily well in
    $\mathcal{H}_{x,j}$ 
    w.r.t. the norm $\| \cdot \|_{\mathcal{H}_j}$.
    By closedness of $\mathcal{H}_{x,j}$ the inclusion holds. 
    By assumption and by definition, respectively, we have that
    $\int_0^T \| k_j(\cdot, x(t))\|_{\mathcal{H}_j}^2  dt < \infty, \quad \|a\|_{L^2} < \infty$.
    Hence, by \cite[Def. 1.2.1]{Hytonen2016BanachSpacesI} it is guaranteed that there is are sequence of simple functions $f_n :  [0,T] \to \mathcal{H}_{x,j}$ and $a_n: [0,T] \to \mathbb{R}$ converging to $k_j(\cdot, x(t))$ and $a(t)$ in $L^2([0,T]; \mathcal{H}_{x,j})$ and $L^2([0,T]; \mathbb{R})$, respectively. 
    By expansion with $\pm k_j(\cdot, x(t))a_n(t)$, triangle inequality and Cauchy-Schwarz, one concludes that
    \setlength{\abovedisplayskip}{2pt}
    \setlength{\belowdisplayskip}{2pt}
    \begin{align*}
         & \int_0^T \| k_j(\cdot, x(t)) a(t) - f_n(t)a_n(t)\|_{\mathcal{H}_j} dt \underset{n \to \infty}{\longrightarrow} 0
    \end{align*}
    Notice, that $(a_nf_n)_{n=1}^\infty$ is a sequence of simple function in $L^1([0,T], \mathcal{H}_{x,j})$. Hence, $\int_0^Tf_n(t)a_n(t)dt$ is a linear combination of elements of $\mathcal{H}_{x,j}$ (which is again in $\mathcal{H}_{x,j}$). \\
    Next, we conclude, $\tilde{\mathcal{H}}_{x,j} = \overline{\tilde{\mathcal{H}}_{x,j, pre}} \subset \overline{\mathcal{H}_{x,j}} = \mathcal{H}_{x,j}$. \\
    Now, we show that the inclusion is tight. Assume that $\tilde{\mathcal{H}}_{x,j, pre}$ is not dense in $\mathcal{H}_{x,j}$. Then there is an element $0 \neq f \in \mathcal{H}_{x,j}$ such that $\langle f, \tilde{\mathcal{H}}_{x,j, pre}\rangle = 0$. Hence, $ \forall a \in L^2$ 
    \setlength{\abovedisplayskip}{2pt}
    \setlength{\belowdisplayskip}{2pt}
    \begin{align*}
        0 & = \left\langle f , \int_0^T k_j( \cdot , x(t))  a(t) dt \right\rangle_{\mathcal{H}_j} =\int_0^T f(x(t)) a(t) dt.
    \end{align*}
    With $a(t) = sign(f(x(t)))$, we conclude $f(x(\cdot)) = 0$ a.e.
    By continuity: $f(x(s)) = \langle f, k_j(\cdot, x(s))\rangle_{\mathcal{H}_j} = \langle f, k_j(\cdot, \lim_i x(s_i))\rangle_{\mathcal{H}_j}  = \lim_i \langle f, k(\cdot, x(s_i))\rangle_{\mathcal{H}_j} = \lim_i f(x(s_i))$, it holds that $f(x(\cdot)) = 0$ on $[0,T]$. But that means that $f \perp k_j(\cdot, x) \ \forall x \in x([0,T])$. That implies $f \in H_{x,j} \cap H_{x,j}^\perp$. But $f = 0$ is a contradiction.
\end{proof}
\begin{proof} (Prop.~\ref{thm:adjoint_operator})
    For \(a\in L_2([0,T])\) and \(f\in \mathcal{H}_j\),
    \begin{align*}
        \langle G_{x,j}a,f\rangle_{\mathcal{H}_j} &= \left\langle \int_0^T k_j(\cdot,x(t))a(t)\,dt, f \right\rangle_{\mathcal{H}_j} \\
        & = \langle f(x(\cdot)),a\rangle_{L_2} =  \langle G_{x,j}^*f,a\rangle_{L_2}
    \end{align*}
\end{proof}
\begin{proof} (Prop.~\ref{thm:G_compact})
    First observe that $G_{x,j}^*$ is bounded:
    \setlength{\abovedisplayskip}{2pt}
    \setlength{\belowdisplayskip}{2pt}
    \begin{align*}
        \|G_{x,j}^*f\|_{L_2}^2 &= \!\! \int_0^T\!\! |f(x(s))|^2\,ds =\!\! \int_0^T \!\!|\langle f,k_j(\cdot,x(s))\rangle_{\mathcal{H}_j}|^2\,ds \\[-5pt]
        & \le \|f\|_{\mathcal{H}_j}^2 \int_0^T k_j(x(s),x(s))\,ds.
    \end{align*}
    Now let \((f_n)\subset \mathcal{H}_j\) be bounded by \(\|f_n\|_{\mathcal{H}_j}\le 1\). Then 
    \[ |G_{x,j}^*f_n(s)| = |f_n(x(s))| \le 1 \cdot \sqrt{k_j(x(s),x(s))} \]
    so the sequence is pointwise bounded. Moreover, 
    \begin{align*}
        &|G_{x,j}^*f_n(s)-G_{x,j}^*f_n(t)| = |f_n(x(s))-f_n(x(t))| \\
        & \le 1 \cdot\|k_j(\cdot,x(s))-k_j(\cdot,x(t))\|_{\mathcal{H}_j} \\
        &= \sqrt{ k_j(x(s),x(s)) + k_j(x(t),x(t)) - 2k_j(x(s),x(t)) }.
    \end{align*}
    Since \(k_j\) is continuous, the right-hand side is uniformly continuous on the compact set \([0,T]^2\). Therefore \((G_{x,j}^*f_n)\) is equicontinuous. Arzelà--Ascoli yields a uniformly convergent subsequence $f_{n_m}$. Since the embedding 
    \( C([0,T]) \hookrightarrow L_2([0,T]) \)
    is linear and continuous/bounded, this subsequence converges in \(L_2\). Hence \(G_{x,j}^*\) is compact. The compactness of \(G_{x,j}\) follows from~\cite[Thm. 4.12]{Kress2014} and as a composition of compact operators, $G_{x,j}^*G_{x,j}$ is compact. 
\end{proof} 
\begin{proof} (Thm.~\ref{thm:time_rescaling}) (cf.~\cite{chicone2024ordinary})
Firstly, notice that $d$ is strictly monotonously increasing in~\eqref{eq:time_rescaling} and~\eqref{eq:original_time_rescaling}, respectively. Hence, it is invertible in both cases. \\
$\implies$: ~\eqref{eq:original_time_rescaling} holds by definition and~\eqref{eq:original_dynamics} by chain rule. \\
$\impliedby$: ~\eqref{eq:time_rescaling} holds by definition and~\eqref{eq:rescaled_gamma} by chain rule. 
\end{proof}
\begin{proof} (Thm.~\ref{thm:derivative_objective}) It suffices to show that the derivatives of the summands
    \setlength{\abovedisplayskip}{2pt}
    \setlength{\belowdisplayskip}{2pt}
    \begin{equation*}
        J_{ij}(x, a_{x,j,\alpha}) = k_j(x_i , x_i)  - \int_0^M k_j(x_i, x(t))a_{x,j,\alpha}(t;x_i)dt
    \end{equation*}
    in the objective~\eqref{eq:rescaled_objective} are
    \setlength{\abovedisplayskip}{2pt}
    \setlength{\belowdisplayskip}{2pt}
    \begin{equation*}
    \begin{split}
        &DJ_{ij}(x,a))(h,\xi) \\
        &= -\int_0^M \partial_2 k_j(x_i, x(t))h(t)a_{x,j,\alpha}(t) + k_j(x_i, x(t))\xi(t) dt.
    \end{split}
    \end{equation*}
    Since $x_i$ is fixed, the term $k_j(x_i,x_i)$ has zero derivative. Because of that, it suffices to consider just the integral term:
    \begin{equation*} 
    \begin{split} & R_{ij}(h,\xi) \\
    &:=\! J_{ij}(x\! +\! h, a_{x,j,\alpha} \!+ \!\xi)\! -\! J_{ij}(x, a_{x,j,\alpha})\! - \!DJ_{ij}{x,a_{x,j,\alpha} (h,\xi)} \\
    &\ =\!-\!\int_0^M\! \bigl[ k_j(x_i,x(t)+h(t)) \bigl(a_{x,j,\alpha}(t)+\xi_j(t)\bigr) \\ &\quad \quad -k_j(x_i,x(t))a_{x,j,\alpha}(t) \\ 
    &\quad \quad -\partial_2 k_j(x_i,x(t))h(t) a_{x,j,\alpha}(t)  -k_j(x_i,x(t))\xi_j(t) \bigr]\,dt . \end{split} 
    \end{equation*} 
    Adding and subtracting $\partial_2 k_j(x_i,x(t))h(t)\xi_j(t)$
    in the integrand and triangle inequality yields
    \begin{equation*} 
    \begin{split} |R_{ij}(h,\xi)| \leq{}& \int_0^M \bigl| k_j(x_i,x(t)+h(t)) -k_j(x_i,x(t)) \\ &\ -\partial_2 k_j(x_i,x(t))h(t) \bigr| \bigl( |a_{x,j,\alpha}(t)|+|\xi_j(t)| \bigr)\,dt \\ +& \int_0^M \bigl| \partial_2 k_j(x_i,x(t)) h(t)\xi_j(t) \bigr|\,dt . 
    \end{split} 
    \end{equation*} 
    Since $k_j\in C^2$ and $x([0,M]) \oplus B_\infty(0, \|h\|_{L^\infty})$ is compact, Taylor's theorem in remainder form implies,
    \setlength{\abovedisplayskip}{2pt}
    \setlength{\belowdisplayskip}{2pt}
    \begin{equation*}
        \bigl| k_j(x_i,x(t)\!+\!h(t)\!) \!-\!k_j(x_i,x(t)\!) \!-\!\partial_2 k_j(x_i,x(t)\!)h(t) \! \bigr|  \!\leq\! C_1 \|h\|_{L^\infty}^2.
    \end{equation*}
    Similar, by continuity of $\partial_2 k_j$, we have \(|\partial_2 k_j(x_i,x(t))|\leq C_2 \) on $x([0,M]) \oplus B_\infty(0, \|h\|_{L^\infty})$. Consequently, it holds
    \begin{equation*}
        |R_{ij}(h,\xi)| \! \leq \! C_1\|h\|_\infty^2 \!\left( \|a_{x,j,\alpha}\|_{L^1} \!\!+\!\!\|\xi_j\|_{L^1} \right)\!+\!C_2\|h\|_\infty\|\xi_j\|_{L^1}.
    \end{equation*}
    Since $[0,M]$ has finite measure, 
    \( \|\xi_j\|_{L^1} \leq \sqrt{M}\,\|\xi_j\|_{L^2}. \) 
    Hence, after absorbing the fixed quantities into $C>0$, 
    \setlength{\abovedisplayskip}{2pt}
    \setlength{\belowdisplayskip}{2pt}
    \begin{equation*} \begin{split} |R_{ij}(h,\xi)| &\leq C\bigl( \|h\|_\infty^2 +\|h\|_\infty^2\|\xi_j\|_{L^2} +\|h\|_\infty\|\xi_j\|_{L^2} \bigr). \\
    &\leq C(\|h\|_\infty+\|\xi_j\|_{L^2})^2,\end{split} \end{equation*} 
    as $\|h\|_{L^\infty}, \|\xi_j\|_{L^2} \leq 1$ wlog. Hence, we have
    \[ \frac{|R_{ij}(h,\xi)|} {\|h\|_\infty+\|\xi_j\|_{L^2}} \leq C(\|h\|_\infty+\|\xi_j\|_{L^2}) \xrightarrow{(h,\xi_j)\to(0,0)} 0, \] 
    which proves the asserted derivative. Summing over $i$ and $j$ and multiplying by the $w_i$s gives the derivative of \eqref{eq:rescaled_objective}.
\end{proof}
\begin{proof} (Thm.~\ref{thm:differentiability_of_variational_equation})
The proof is an application of the implicit function theorem in Banach spaces~\cite[Thm. 5.9.]{implicite_function}. Three premises have to be shown: 
1. \( f\) is Fréchet differentiable. The candidate for the derivative at \((x,a)\) is
\begin{equation*}
    \begin{aligned}
        &Dg(x,a)(h,\zeta) = \\[-1pt]
        &\int_0^T\!\! k_1(x(\cdot),x(s)) h(\cdot)a(s)\,ds\! + \!\int_0^T \!\!k_2(x(\cdot),x(s)) h(s)a(s)\,ds \\[-5pt]
        &+ \int_0^T k(x(\cdot),x(s)) \zeta(s)\,ds + \alpha\zeta(\cdot) - k_1(x(\cdot),x)h(\cdot).
    \end{aligned}
\end{equation*}
This operator is indeed the Fréchet derivative: 
    \setlength{\abovedisplayskip}{2pt}
    \setlength{\belowdisplayskip}{2pt}
\begin{align*} 
&\Bigg\| \int_0^T k(x(\cdot)+h(\cdot),x(s)+h(s)) (a(s)+\zeta(s))\ \\
& \quad \! \!\! \! + \alpha(a(\cdot)+\zeta(\cdot))  - k(x(\cdot)+h(\cdot),x) ds \\
& \quad \! \!\! \! - \int_0^T k(x(\cdot),x(s))a(s)\,ds  - \alpha a(\cdot)  + k(x(\cdot),x) \\ 
&\quad \! \!\! \! - \int_0^T\!\!\!\!\! k_1(x(\cdot),x(s))h(\cdot)a(s)\,ds \! -\!\! \!\int_0^T\!\!\!\!\! k_2(x(\cdot),x(s))h(s)a(s)\,\!ds \\
& \quad \! \!\! \!- \int_0^T k(x(\cdot),x(s))\zeta(s)\,ds - \alpha\zeta(\cdot)+ k_1(x(\cdot),x)h(\cdot) \Bigg\|_{L_2}
\end{align*}
\begin{align*}
&\le \int_0^T \Big\| k(x(\cdot)+h(\cdot),x(s)+h(s)) (a(s)+\zeta(s)) \\
& \quad - k_1(x(\cdot),x(s)) h(\cdot)(a(s)+\zeta(s)) - k(x(\cdot),x(s)) \cdot  \\ 
&\quad \cdot (a(s)+\zeta(s)) - k_2(x(\cdot),x(s)) h(s)(a(s)+\zeta(s)) \Big\|_{L_2} \,ds \\[-5pt]
&\quad + \int_0^T \Big\| k_1(x(\cdot),x(s)) h(\cdot)(a(s)+\zeta(s)) \\
& \quad + k_2(x(\cdot),x(s)) h(s)(a(s)+\zeta(s)) \\ 
&\quad - k_1(x(\cdot),x(s)) h(\cdot)a(s) - k_2(x(\cdot),x(s)) h(s)a(s) \Big\|_{L_2} \,ds \\ 
&\quad +0 + \Big\| k(x(\cdot)\!+\!h(\cdot),x) \!-\! k(x(\cdot),x) \!-\! k_1(x(\cdot),x)h(\cdot) \Big\|_{L_2} \\
&\le \int_0^T \Big\| C_1 \bigl( |h(\cdot)|^2+h(s)^2 \bigr) |a(s)+\zeta(s)| \Big\|_{L_2} \,ds \\
&\quad + \int_0^T \Big( \| k_1(x(\cdot),x(s)) h(\cdot)\zeta(s) \|_{L_2}  \\
& \quad + \| k_2(x(\cdot),x(s)) h(s)\zeta(s) \|_{L_2} \Big)\,ds + \|C_2 h(\cdot)^2\|_{L_2} \\[1ex]
&\le C_1 \,2\|h\|_{L_\infty}^2 \bigl( \|a\|_{L_1} + \|\zeta\|_{L_1} \bigr) \\
& \quad + \|h\|_{L_\infty} C_3 \,2\|\zeta\|_{L_1} + C_2\|h\|_{L_\infty}^2 \\
&\le M \Big( \|h\|_{L_\infty}^2 + 2\|h\|_{L_\infty}\|\zeta\|_{L_2} \Big) \\
&\le M \Big( \|h\|_{L_\infty} + \|\zeta\|_{L_2} \Big)^2 \in o \Big( \|h\|_{L_\infty} + \|\zeta\|_{L_2} \Big). 
\end{align*}
Thus \(f\) is Fréchet differentiable. \\
2. Linearity of \(Dg(x,a)\) is immediate. \\
3. The partial derivative with respect to \(a\), 
\[ D g(x,a)(0, \zeta) = \int_0^T k(x(\cdot),x(s)) \zeta(s)\,ds + \alpha\zeta(\cdot), \]
is an isomorphism: Similar to~\eqref{eq:variational_equation}, this operator a compact perturbation of the identity with a positive definite perturbation and hence invertible~\cite[Thm. 3.4]{Kress2014}. 
Now the claim follows directly from the implicit function theorem.
\end{proof}

\begin{theorem}
    \label{thm:variational_dynamics}
    \cite[Thm. 6.1.1. Proof 2.]{bressan2007introduction}
    Let the system dynamics $\tilde{f}: [0, M] \times \mathcal{X} \times \mathcal{U} \to \mathbb{R}^n$ be continuous and continuously differentiable in the state variable and $u^*$ be a bounded, admissible control. Define the needle variation of the control at $\hat{t} \in (0,M]$ as $u_\varepsilon(t) = \{ \omega, \ \text{if} \ t\in[\hat t-\varepsilon, ], u^*(t), \text{else}\}$.
Define \(v\) by
\begin{equation}
     \label{eq:variational_dynamics_statement_2}
     \begin{aligned}
         \dot v(t) &= \tilde{f}_x(t, x^*(t),u^*(t))v(t),\\  
     v(\hat t)& = \tilde{f}(\hat{t}, x^*(\hat t),\omega) - \tilde{f}(\hat t, x^*(\hat t),u^*(\hat t)). 
     \end{aligned}
\end{equation}
on $[\hat{t}, 1]$ and $v \equiv 0$ on $[0, \hat{t})$.
The corresponding trajectory variation satisfies
\begin{equation}
     \lim_{\varepsilon\to 0} \left\| (x(\cdot;u_\varepsilon)-x^*)/\varepsilon - v \right\|_{L_\infty} = 0.
\end{equation}
\end{theorem}

\begin{proof} (Thm.~\ref{thm:PMP}). By optimality, it holds
\setlength{\abovedisplayskip}{2pt}
    \setlength{\belowdisplayskip}{2pt}
\begin{equation}
\label{eq:optimality}
    \begin{aligned}
        \sum_{i = 1}^N \!w_i\!\sum_{j =1}^n\! \ & k_j(x_i,x_i)\! -\!\! \int_0^M \!\!\!\!k_j(x_i,x(t,u_\varepsilon)) g_{ij}(x(\cdot,u_\varepsilon))(t)\,dt \\
    &- k_j(x_i,x_i) + \int_0^M k_j(x_i,x^*(t)) a_{ij}^*(t)\,dt \quad \geq 0.        
    \end{aligned}
\end{equation}

Step 1: We want to show that
\setlength{\abovedisplayskip}{2pt}
    \setlength{\belowdisplayskip}{2pt}
\begin{equation}
\label{eq:gradient_optimality}
\begin{aligned}
    \sum_{i = 1}^Nw_i\sum_{j =1}^n -\int_0^M & \Big( \partial_2 k_j(x_i,x^*(t)) v(t)a_{ij}^*(t) \\
    &+ k(x_i,x^*(t)) D_{x^*,a_{ij}^*}g(v)(t) \Big) \,dt \geq 0. \!\!\!\!\!\!\!\!\!\!\!\!
\end{aligned}
\end{equation}
We argue by contradiction. Assume there exists $\hat t\in (0,T]$ and $\omega\in \mathcal{U}$ and an $M > 0$ such that 
\begin{align*}
    \sum_{i = 1}^Nw_i\sum_{j =1}^n -\int_0^M & \Big( \partial_2 k_j(x_i,x^*(t)) v(t)a_{ij}^*(t) \\[-5pt]
    & \!\!\!+ k(x_i,x^*(t)) D_{x^*,a_{ij}^*}g_{ij}(v)(t) \Big) \,dt < -M. 
\end{align*}
We observe that the functional 
\setlength{\abovedisplayskip}{2pt}
    \setlength{\belowdisplayskip}{2pt}
\begin{equation}
    \label{eq:derivative_functional}
    \begin{split}
        \hat{v} \mapsto  \sum_{i = 1}^Nw_i\sum_{j =1}^n &-\int_0^M \Big( \partial_2 k_j(x_i,x^*(t)) \hat{v}(t)a_{ij}^*(t) \\[-5pt]
    &+ k(x_i,x^*(t)) D_{x^*,a_{ij}^*}g_{ij}(\hat{v})(t) \Big) \,dt
    \end{split}
\end{equation}
i.e. from $L^\infty$ to $\mathbb{R}$ is linear (trivial) and bounded: 
\setlength{\abovedisplayskip}{2pt}
    \setlength{\belowdisplayskip}{2pt}
\begin{align*}
    &\Bigg|\sum_{i = 1}^Nw_i\sum_{j =1}^n -\int_0^M \Big( \partial_2 k_j(x_i,x^*(t)) \hat{v}(t)a_{ij}^*(t)\\[-7pt]
    &\qquad \qquad + k(x_i,x^*(t)) D_{x^*,a_{ij}^*}g_{ij}(\hat{v})(t) \Big) \,dt\Bigg| \\[-7pt]
    &\leq \Bigg|\sum_{i = 1}^Nw_i\sum_{j =1}^n \int_0^M |\partial_2 k_j(x_i,x^*(t))||\hat{v}(t)||a_{ij}^*(t)|\\[-7pt]
    &\qquad \qquad  + |k(x_i,x^*(t))| |(D_{x^*,a_{ij}^*}g_{ij} (\hat{v}) (t)| \Big) \,dt\Bigg| \\[-7pt]
     &\leq \Bigg|\sum_{i = 1}^Nw_i\sum_{j =1}^n \|\partial_2 k_j(x_i,x^*(t))\|_{L^2}\|\hat{v}\|_{L^\infty}\|a_{ij}^*\|_{L^2}\\[-4pt]
    &\qquad \qquad + \|k(x_i,x^*(t))\|_{L^2} \|D_{x^*,a_{ij}^*}g_{ij}\| \|\hat{v}\|_{L^2}  \leq C \|\hat{v}\|_{L^\infty}
\end{align*}
The last inequality followed from $k_j$ being continuously differentiable and from $M < \infty$.  From theorem~\ref{thm:variational_dynamics} we know that:
\setlength{\abovedisplayskip}{2pt}
    \setlength{\belowdisplayskip}{2pt}
\begin{equation}
    \label{eq:variation_statement}
    \lim_{\varepsilon\to0} \left\|(x(\cdot,u_\varepsilon)-x^*(\cdot))/\varepsilon-v(\cdot)\right\|_{L_\infty}=0.
\end{equation}
By continuity of the functional~\eqref{eq:derivative_functional} we get that there exists a \(\delta>0\) such that
\setlength{\abovedisplayskip}{2pt}
    \setlength{\belowdisplayskip}{2pt}
\begin{equation}
    \label{eq:contradition_3}
    \begin{aligned}
        - \sum_{i = 1}^Nw_i\sum_{j =1}^n & \int_0^M \Big( \partial_2 k_j(x_i,x^*(t)) y(t)a_{ij}^*(t) \\[-5pt]
    &+ k(x_i,x^*(t)) D_{x^*,a_{ij}^*}g_{ij}(y)(t) \Big) \,dt < -\frac{1}{2}M
    \end{aligned} \!\!\!\!\!\!\!\!
\end{equation}
for all \(y\) satisfying $\|y\!-\!v\|_{L_\infty} \!\! < \!\!\delta$. In particular, it holds for
\setlength{\abovedisplayskip}{2pt}
    \setlength{\belowdisplayskip}{2pt}
\begin{equation}
    \label{eq:inserting_variation}
    y = (x(\cdot,u_{\varepsilon(\delta)})-x^*(\cdot))/\varepsilon(\delta).
\end{equation}
For the next equivalence statement, we define
\setlength{\abovedisplayskip}{2pt}
    \setlength{\belowdisplayskip}{2pt}
\begin{align*}
    A_{ij} := & k_j(x_i, x(t, u_\varepsilon) -  k_j(x_i, x^*(t)) \\
     &- \partial_2k_j(x_i, x^*(t)) (x(t,u_\varepsilon) - x^*(t)) \\
    B_{ij} := & g_{ij}(x(\cdot, u_\varepsilon))(t)\! -\! g_{ij}(x^*)(t)\! -\! Dg_{ij}(x(\cdot, u_\varepsilon) - x^*)(t).
\end{align*}
Inserting~\eqref{eq:inserting_variation} into~\eqref{eq:contradition_3} gives
\setlength{\abovedisplayskip}{2pt}
    \setlength{\belowdisplayskip}{2pt}
\begin{align*}
    & - \sum_{i = 1}^Nw_i\sum_{j =1}^n \int_0^M \big(A_{ij}\cdot B_{ij} - k_j(x_i, x(t,u_\varepsilon)) \cdot B_{ij} \\
    & \quad - g_{ij}(x(\cdot, u_\varepsilon))(t)\cdot A_{ij} +  k_j(x_i, x(t,u_\varepsilon)) g_{ij}(x(\cdot, u_\varepsilon))(t) \\
    & \quad - k_j(x_i, x^*(t)) g_{ij}(x^*)(t)\big)/\varepsilon(\delta) dt \qquad\qquad < -\frac{1}{2}M.
\end{align*}
This can be checked be multiplying out the terms involving $A_{ij}$ and $B_{ij}$. 
With~\eqref{eq:optimality} it follows:
\setlength{\abovedisplayskip}{2pt}
    \setlength{\belowdisplayskip}{2pt}
\begin{align*}
    & - \sum_{i = 1}^Nw_i\sum_{j =1}^n \int_0^M \big(A_{ij}\cdot B_{ij} - k_j(x_i, x(t,u_\varepsilon)) \cdot B_{ij} \\[-5pt]
    & \quad - g_{ij}(x(\cdot, u_\varepsilon))(t)\cdot A_{ij} \big)/\varepsilon(\delta) dt \qquad\qquad < -\frac{1}{2}M.
\end{align*}
Taking norms, triangle inequalities, and Cauchy-Schwarz implies:
\setlength{\abovedisplayskip}{2pt}
    \setlength{\belowdisplayskip}{2pt}
\begin{align*}
    &  \sum_{i = 1}^N \! w_i \! \sum_{j =1}^n C \|A_{ij}\|_{L^2} \| B_{ij} \|_{L^2} \! + \!\|k_j(x_i, x(\cdot,u_\varepsilon))\|_{L^2} \| B_{ij}\|_{L^2} \\[-5pt]
    & \quad \| g_{ij}(x(\cdot, u_\varepsilon))(\cdot)\|_{L^2} \|A_{ij}\|_{L^2} \big)/\varepsilon(\delta) dt \qquad\qquad > \frac{1}{2}M.
\end{align*}
We can bound $\|k_j(x_i, x(\cdot,u_\varepsilon))\|_{L^2}$ uniformly for small enough $\varepsilon$ as $k_j$ is continuous and the possible second arguments are on a compact domain and we can bound $ \| g_{ij}(x(\cdot, u_\varepsilon))(\cdot)\|_{L^2}$ for small enough $\varepsilon$ as the map $g_{ij}$ is continuous. Then we expand each $A_{ij}$ and $B_{ij}$ with $\|x(\cdot, u_\varepsilon)) - x^*\|_{L^\infty}$, what gives:
\setlength{\abovedisplayskip}{2pt}
    \setlength{\belowdisplayskip}{2pt}
\begin{align*}
    &  \sum_{i = 1}^Nw_iC\sum_{j =1}^n  \frac{\|A_{ij}\|_{L^2}}{\|x(\cdot, u_\varepsilon)) - x^*\|_{L^\infty}} \frac{\|B_{ij}\|_{L^2}}{\|x(\cdot, u_\varepsilon)) - x^*\|_{L^\infty}} \cdot \\
    & \cdot \frac{\|x(\cdot, u_\varepsilon)) - x^*\|_{L^\infty}}{\varepsilon(\delta)}\cdot\|x(\cdot, u_\varepsilon)) - x^*\|_{L^\infty} \\
    &+ \frac{\|B_{ij}\|_{L^2}}{\|x(\cdot, u_\varepsilon)) - x^*\|_{L^\infty}}\frac{\|x(\cdot, u_\varepsilon)) - x^*\|_{L^\infty}}{\varepsilon(\delta)} \\
    & + \frac{\|A_{ij}\|_{L^2}}{\|x(\cdot, u_\varepsilon)) - x^*\|_{L^\infty}}\frac{\|x(\cdot, u_\varepsilon)) - x^*\|_{L^\infty}}{\varepsilon(\delta)} \qquad\quad > \frac{1}{2}M.
\end{align*}
But for $\varepsilon \to 0$ (\eqref{eq:variation_statement} still guarantees~\eqref{eq:contradition_3}), we get
\setlength{\abovedisplayskip}{2pt}
    \setlength{\belowdisplayskip}{2pt}
\begin{align*}
     &\frac{\|A_{ij}\|_{L^2}}{\|x(\cdot, u_\varepsilon)) \!-\! x^*\|_{L^\infty}}  \!\xrightarrow{\varepsilon \to 0} \!0, && \!\!\!\!\!\!\!
     \frac{\|B_{ij}\|_{L^2}}{\|x(\cdot, u_\varepsilon)) \!-\! x^*\|_{L^\infty}}  \!\xrightarrow{\varepsilon \to 0}\! 0, \\
     &\frac{\|x(\cdot, u_\varepsilon)) \!-\! x^*\|_{L^\infty}}{\varepsilon(\delta)}   \!\!\xrightarrow{\varepsilon \to 0}\! \|v\|_{L^\infty}, && \!\!\!\!\!\!
     \|x(\cdot, u_\varepsilon))\! -\! x^*\|_{L^\infty} \! \xrightarrow{\varepsilon \to 0} \!0,
\end{align*}
which is a contradiction. \\
Step 2: From~\eqref{eq:gradient_optimality}, we conclude that
\setlength{\abovedisplayskip}{2pt}
    \setlength{\belowdisplayskip}{2pt}
\begin{align*}
    0 \leq & \sum_{i=1}^{N} \omega_i \sum_{j=1}^{m} - \int_{\Omega}  k_{2j}^{\,i}\bigl(x_i,x^*(t)\bigr) v(t)\,a_j^*(t) \\[-5pt]
    & \qquad\qquad \qquad + k_j\bigl(x_i,x^*(t)\bigr) D_{x^*}a_j^*\,g_j(v)(t) dt \\[-5pt]
    &+ \int_0^Mq(t)^{T} \underbrace{\left(\dot{v}(t) - \tilde{f}_x\bigl(t,x^*(t),u^*(t)\bigr)v(t) \right)}_{= 0}\,dt. 
\end{align*}
Integration by parts yields: 
\setlength{\abovedisplayskip}{2pt}
    \setlength{\belowdisplayskip}{2pt}
\begin{align*}
    \int_0^M \!\!\!\! q(t)^T\dot v(t)\,dt= &- \int_0^M \dot q(t)^Tv(t)\,dt + q(M)^Tv(M) \\
    &  -\! q(\hat{t}_+)^T\!v(\hat{t}_+)\!+\! q(\hat{t}_-)^T\!v(\hat{t}_-)\! - \!q(0)^T\!v(0) \\
    = &- \int_0^T \dot q(t)^Tv(t)\,dt - q(\hat{t}_+)^Tv(\hat{t}_+),
\end{align*}
since  $q(M)  = v(\hat{t}_-) = v(0) =0$. Plugging in~\eqref{eq:adjoint_equation} yields
\setlength{\abovedisplayskip}{2pt}
    \setlength{\belowdisplayskip}{2pt}
\begin{align*}
    0 \leq &  - q(\hat{t}_+)^Tv(\hat{t}_+) \\
     = &- q(\hat t)^T \Big( \tilde{f}(\hat{t}, x^*(\hat t),\omega) - \tilde{f}(\hat{t}, x^*(\hat t),u^*(\hat t)) \Big)
\end{align*}
Rearranging gives the desired maximum principle:
\setlength{\abovedisplayskip}{2pt}
    \setlength{\belowdisplayskip}{2pt}
\begin{equation*}\qedhere
    q(\hat t)^T \tilde{f}(\hat{t}, x^*(\hat t),\omega) \le q(\hat t)^T \tilde{f}(\hat{t}, x^*(\hat t),u^*(\hat{t}  )) .
\end{equation*}
\end{proof}
\vspace{12pt}

\end{document}